\documentclass[cernpreprint, atlasdraft=false, texlive=2023, USenglish, texmf, orcidlogo]{atlasdoc}
\DeclareOldFontCommand{\it}{\normalfont\itshape}{\mathit}
 
\usepackage{atlaspackage}
\usepackage{atlasbiblatex}
 
\usepackage{atlasphysics}

\graphicspath{{logos/}{figures/}}
 
\usepackage{ANA-EXOT-2018-35-PAPER-defs}

\AtlasTitle{Search for the charged-lepton-flavor-violating decay $Z\rightarrow e\mu$ in $pp$ collisions at $\sqrt{s}=13$~TeV with the ATLAS detector}

\AtlasAbstract{
A search for the charged-lepton-flavor-violating process $Z\rightarrow e\mu$ is presented, using 139~fb$^{-1}$ of $\sqrt{s}=13$~TeV $pp$ collision data collected by the ATLAS experiment at the LHC. An excess in the $e\mu$ invariant mass spectrum near the $Z$ boson mass would be a striking signature of new physics.  No excess is observed, and an upper limit ${\cal B}(Z\rightarrow e\mu) < 2.62 \times 10^{-7}$ is placed on the branching fraction at 95\% confidence level,
which is the most stringent limit to date.
}

\AtlasRefCode{EXOT-2018-35}
 
\PreprintIdNumber{CERN-EP-2022-007}

\AtlasJournalRef{Phys.\ Rev.\ D.\ 108\ (2023)\ 032015}
\AtlasDOI{10.1103/PhysRevD.108.032015}

\hypersetup{pdftitle={ATLAS document},pdfauthor={The ATLAS Collaboration}}
 
\begin{document}
 
\maketitle

\section{Introduction}
 
The observed conservation of charged-lepton flavor is a long-standing mystery. Despite the lack of protection from a fundamental symmetry, no charged-lepton-flavor-violating decays have been observed~\cite{doi:10.1146/annurev-nucl-102912-144530,HIGG-2017-08,HIGG-2018-58,CMS-HIG-20-009}.
 
Lepton-flavor violation has been observed in the neutrino sector~\cite{Super-Kamiokande:2001ljr,SNO:2001kpb}, but the rate of charged-lepton-flavor transitions mediated by neutrino-flavor oscillations is expected to be vanishingly small~\cite{PhysRevD.63.053004}, giving, for example\footnote{
The electric charges of the lepton pairs throughout the paper are omitted for brevity, but opposite charges are implied except when specified.}, $\mathcal{B}$($Z\rightarrow e\mu) <4 \times 10^{-60}$.   New sources of charged-lepton-flavor violation would indicate physics beyond the Standard Model (BSM), and searches for such violations can be used to constrain BSM theories~\cite{Yang:2010iq,Ding:2007jr,Gripaios:2009dq,Hirsch:2009ee}.
 
A search for muon decays into $e^+e^-e^+$ by SINDRUM~\cite{Bellgardt:1987du} and a search for $\mu \rightarrow e\gamma$ by MEG~\cite{TheMEG:2016wtm} imply  ${\cal B}(Z\rightarrow e\mu) < 5 \times 10^{-13}$ at 90\% confidence level (CL)~\cite{PhysRevD.63.016003}.
However, these interpretations are indirect, and can be evaded in intriguing scenarios, such as anomalous magnetic moments or delicate cancellations~\cite{PhysRevD.63.016003},  which cannot be ruled out.  Direct searches for two-body decays into $e\mu$ therefore remain a vital part of the investigation into charged-lepton-flavor violation.
Searches at LEP give ${\cal B}(Z\rightarrow e\mu) < 1.7 \times 10^{-6}$ at 95\% CL~\cite{Akers:1995gz,Abreu:1996mj,Adriani:1993sy,Decamp:1991uy} and a previous search at the Large Hadron Collider yielded ${\cal B}(Z\rightarrow e\mu) < 7.5 \times 10^{-7}$ at 95\% CL~\cite{Aad:2014bca}, in 20.3~fb$^{-1}$ of 8~\TeV\ proton collision data collected by the ATLAS experiment. Searches for $Z\rightarrow \tau\ell$, where $\ell=e$ or $\mu$, report limits of ${\cal B}(Z\rightarrow e\tau) < 5.0 \times 10^{-5}$ and ${\cal B}(Z\rightarrow \mu\tau) < 6.5 \times 10^{-6}$ at 95\% CL~\cite{ATLAS:2021bdj}.
 
This paper presents a search for $Z\rightarrow e\mu$ using 139~fb$^{-1}$ of proton collision data collected at $\sqrt{s}=13$~\TeV, in which a boosted decision tree and a veto on $b$-quark-tagged jets are used to enhance the signal selection.
 
\section{ATLAS detector}
 
The ATLAS detector~\cite{PERF-2007-01} consists of an inner detector (ID) surrounded by a solenoid that produces a 2~T magnetic field, electromagnetic (EM) and hadronic calorimeters, and a muon spectrometer (MS) immersed in a magnetic field produced by a system of toroidal magnets. The ID measures the trajectories of charged particles over the full azimuthal angle and in the pseudorapidity\footnote{
ATLAS uses a right-handed coordinate system with its origin at the nominal interaction point (IP) in the center of the detector and the $z$-axis along the beam pipe.
The $x$-axis points from the IP to the center of the LHC ring, and the $y$-axis points upward. Cylindrical coordinates $(r,\phi)$ are used in the transverse plane, $\phi$ being the azimuthal angle around the beam pipe.
The pseudorapidity is defined in terms of the polar angle $\theta$ as $\eta = -\ln\tan(\theta/2)$.
Transverse momentum and energy are defined relative to the beamline as
$p_{\textrm{T}} = p \sin(\theta)$ and $E_{\textrm{T}} = E \sin(\theta)$.
}
range of $|\eta| < 2.5$ using silicon pixel, silicon microstrip, and straw-tube transition-radiation tracker detectors. Liquid-argon (LAr) EM sampling calorimeters cover the range $|\eta| < 3.2$ and a scintillator-tile calorimeter provides hadronic calorimetry for $|\eta| < 1.7$. In the endcaps ($|\eta| > 1.5$), LAr is also used for the hadronic calorimeters, matching the outer $|\eta|$ limit of endcap EM calorimeters. The LAr forward calorimeters extend the coverage to $|\eta| < 4.9$ and provide both the EM and hadronic energy measurements. The MS measures the deflection of muons within $|\eta| < 2.7$ using three stations of precision drift tubes, with cathode strip chambers in the innermost station for $|\eta| > 2.0$, and provides separate trigger measurements from dedicated chambers in the region $|\eta| < 2.4$.
A trigger system implemented with hardware and software components is used to select interesting events to be recorded for subsequent offline analysis~\cite{TRIG-2016-01}.
An extensive software suite~\cite{ATL-SOFT-PUB-2021-001} 
is used in the reconstruction and analysis of real and simulated data, in detector operations, and in the trigger and data acquisition systems of the experiment.

\section{Analysis strategy}
 
The search for flavor-violating decays of the $Z$ boson is performed by examining the invariant mass distribution of opposite-charge $e\mu$ candidates for evidence of a narrow peak consistent with direct $Z$ boson decay.
The event selection requires two isolated energetic, oppositely charged leptons of different flavor: $e^\pm\mu^\mp$.  The primary backgrounds consist of decays into $\tau$-lepton pairs ($Z \to \tau \tau \to e \mu \nu \bar{\nu} \nu \bar{\nu}$),
decays into muon pairs $Z \to \mu \mu$ where one muon is misidentified as an electron, dileptonic final states from decays of top quark pairs ($t\bar{t}\to e\mu \nu \bar{\nu} b\bar{b}$), and  decays of weak boson pairs ($WW\to e \mu \nu \bar{\nu}$).
To suppress the contribution from top quark and boson pairs,  events are required to have little jet activity and only a small amount of missing transverse momentum.
To further reduce the background, a multivariate boosted decision tree (BDT)~\cite{Hocker:2007ht} is trained to distinguish between signal and background events, and the BDT output must exceed a threshold selected to optimize the ratio of expected signal to the square root of the expected background yield. Events from background processes which pass the selection are expected to form a smooth spectrum in the electron--muon invariant mass ($m_{e\mu}$) within the window $70<m_{e\mu}< 110$~\GeV. A binned likelihood fit, in which the signal is unconstrained, is performed. In the absence of a signal, an upper limit on the branching fraction $\mathcal{B}$($Z\rightarrow e\mu$) is set, related to a ratio of
the observed $e\mu$ yield to the average of the observed yields of $ee$ and $\mu\mu$ events to cancel out common systematic uncertainties.

\section{Monte Carlo samples}
 
Samples of simulated collisions generated using Monte Carlo (MC) methods are used to estimate the dominant backgrounds as well as to optimize the event selection.  All MC samples were produced using the ATLAS detector simulation~\cite{Aad:2010ah} based on $\GEANT4$~\cite{Agostinelli:2002hh}.  Simulated signal $Z\rightarrow e\mu$ events were generated at leading order with $\Pythia\,8.210$~\cite{Sjostrand:2014zea} using the A14 set of tuned parameters (tune)~\cite{ATL-PHYS-PUB-2014-021} and the \NNPDF[2.3lo] parton distribution function (PDF) set~\cite{Ball:2012cx}.
 
Background events with leptonically decaying $W$ bosons or $Z\rightarrow\tau\tau$ production in association with jets were simulated with the
$\SHERPA\,2.2.1$~\cite{Bothmann:2019yzt}
generator using next-to-leading-order (NLO) matrix elements for up to two partons, and leading-order (LO) matrix elements
for up to four partons, calculated with the \textsc{Comix}~\cite{Gleisberg:2008fv}
and \OPENLOOPS~\cite{Buccioni:2019sur,Cascioli:2011va,Denner:2016kdg} libraries. They
were matched with the \SHERPA parton shower~\cite{Schumann:2007mg} using the \MEPSatNLO
prescription~\cite{Hoeche:2011fd,Hoeche:2012yf,Catani:2001cc,Hoeche:2009rj}
and the set of tuned parameters developed by the \SHERPA authors.
The \NNPDF[3.0nnlo] set of PDFs~\cite{Ball:2014uwa} was used and the samples
were normalized to a cross-section prediction at next-to-next-to-leading order (NNLO) in QCD~\cite{Anastasiou:2003ds}.
 
Background events with $Z\rightarrow \mu\mu$ or $Z\rightarrow ee$ in association with jets were modeled with
the \textsc{Powheg\,Box}\,v1 MC generator~\cite{Nason:2004rx,Frixione:2007vw,Alioli:2010xd,Alioli:2008gx}
at NLO in the hard-scattering processes of $Z$ boson production. It was interfaced to $\PYTHIA\,8.186$~\cite{Sjostrand:2007gs}
for the modeling of the parton shower, hadronization, and underlying
event, with parameters set according to the \AZNLO
tune~\cite{STDM-2012-23}. The \CT[10nlo] PDF set~\cite{Lai:2010vv} was used
for the hard-scattering processes, whereas the CTEQ6L1 PDF
set~\cite{Pumplin:2002vw} was used for the parton shower. The effect of
QED final-state radiation was simulated with \textsc{Photos++}\,3.52~\cite{Golonka:2005pn,Davidson:2010ew}.
The \textsc{EvtGen}\,1.2.0 program~\cite{Lange:2001uf} was used to decay bottom and charm hadrons.

All the $Z$ boson samples, including $Z\rightarrow \tau\tau$, $Z\rightarrow \mu\mu$, $Z\rightarrow ee$ and $Z\rightarrow e\mu$ events, were reweighted such that the transverse momentum ($p_\textrm{T}$) of the $Z$ boson matches that observed in $Z\rightarrow ee$ and $Z\rightarrow \mu\mu$ decays in data~\cite{ATLAS:2019zci}.
 
Samples of events with fully leptonic diboson final states
and semileptonic diboson final states, where one boson
decays leptonically and the other hadronically, were simulated with the
$\SHERPA\,2.2.1$ or 2.2.2~\cite{Bothmann:2019yzt} generator depending on the diboson ($VV$) process,
including off-shell effects and Higgs boson contributions where appropriate.
They were generated using
matrix elements at NLO for up to one additional parton
and at LO accuracy for up to three additional parton
emissions. Samples for the loop-induced processes $gg \to VV$ were
generated using LO-accurate matrix elements for up to one
additional parton emission for both the cases of fully leptonic and
semileptonic final states. The matrix element calculations were matched
and merged with the \SHERPA parton shower based on Catani--Seymour
dipole factorization~\cite{Gleisberg:2008fv,Schumann:2007mg} using the MEPS@NLO
prescription.
The virtual QCD corrections were provided by the
\OPENLOOPS library. The
\NNPDF[3.0nnlo] set of PDFs was used, along with the
dedicated set of tuned parton-shower parameters developed by the \SHERPA authors. The cross section for the $WW\to e \mu \nu \bar{\nu}$ processes was normalized to a prediction at NNLO in QCD~\cite{2016WWcrosssection}.

The top quark backgrounds, i.e.\ $t\bar{t}$ and single top production, were modeled with the
\textsc{Powheg\,Box}\,v2~\cite{Frederix:2012dh,Nason:2004rx,Frixione:2007vw,Alioli:2010xd}
generator at NLO, using the four-flavor scheme and the
\NNPDF[3.0nlo] set of PDFs.  The events were
interfaced with $\PYTHIA\,8.230$ using the A14
tune and the \NNPDF[2.3lo] set of PDFs.

Additional $pp$ collisions from the same bunch crossing are included in each event, according to the distribution observed in data.  The simulated events are reconstructed with the same software as the data.
 
\section{Object selection}
 
Events are required to have at least one primary collision vertex that has at least two associated tracks, each with transverse momentum $p_{\textrm{T}} > 0.5$~\GeV.
The primary vertex is selected as the one with the largest \(\Sigma \pT^2\),
where the sum is over all tracks with transverse momentum \(\pT > \SI{0.4}{\GeV}\) that are associated with the vertex.
 
Candidate electrons are required to have $p_{\textrm{T}} > 27$~\GeV\ and pseudorapidity $|\eta| < 1.37$ or \mbox{$1.52 < |\eta| < 2.47$} to ensure they are contained in the high-granularity  region of the EM calorimeter and avoid the transition region between the barrel and endcap calorimeters.  Candidates must satisfy the ``TightLH'' identification requirements~\cite{PERF-2018-01}, which are based on calorimeter shower shape, ID track quality, and the spatial match between the shower and the track, as well as the ``Tight'' isolation requirement~\cite{PERF-2018-01}. The track parameters $z_0$ and $\theta$ are the longitudinal impact parameter and the polar angle of the electron candidate at the point of closest approach of the track to the beam, respectively. Electrons are required to have a transverse impact parameter with respect to the measured beam position of magnitude less than $5\sigma$, where $\sigma$ is its estimated uncertainty, and $|z_0 \sin\theta|$ less than 0.5~mm.
 
Candidate muons are required to have $p_{\textrm{T}} > 27$~\GeV\ and $|\eta| < 2.5$. Candidates must also satisfy the ``Medium'' identification requirements~\cite{MUON-2018-03}, which are based on track quality, as well as the ``Tight'' isolation requirement~\cite{Aad:2016jkr}.  Muons are also required to have  a transverse impact parameter with respect to the measured beam position of magnitude less than $3\sigma$, and $|z_0 \sin\theta|$ less than 0.5~mm.
 
Hadronic jets are reconstructed from topological clusters~\cite{ATLAS:2016krp} of energy deposits in the EM and hadronic calorimeters using the anti-$k_\textrm{t}$ algorithm~\cite{Cacciari:2008gp,Cacciari:2011ma} with distance parameter $R=0.4$.
The topological clusters are calibrated at the EM energy scale. The jets are fully calibrated using the EM\,+\,jet energy scale scheme~\cite{PERF-2016-04}, and required to have $p_{\textrm{T}} >20$~\GeV\ and $|\eta|<2.5$. To reject jets from other $pp$ collisions (pileup), candidate jets with $p_\textrm{T}<60$~\GeV\ and $|\eta|<2.4$ are required to
pass the jet vertex tagger~\cite{Aad:2015ina}, a likelihood discriminant combining information from several track-based variables.
 
Jets containing $b$-hadrons are tagged if they satisfy the requirements of the highest-efficiency working point (85\% for jets in $t\bar{t}$ events containing $b$-hadrons) of the MV2c10 multivariate tagging algorithm~\cite{Aad:2019aic}, which is based on track impact parameters and secondary vertices that are reconstructed from the tracks with large impact parameter significances.
 
The missing transverse momentum (with magnitude $E_\textrm{T}^\textrm{miss}$) is calculated as the negative vectorial sum of the $p_\textrm{T}$ of all reconstructed and calibrated electrons, muons, tau leptons, photons and jets~\cite{ATLAS-CONF-2018-023,Aaboud:2018tkc}, as well as inner-detector tracks originating from the primary vertex but not associated with any reconstructed objects.
 
\section{Event selection}
 
The dataset used in this search was collected during LHC Run~2 in stable beam conditions and with all detector systems operating normally.  For this search, performed in 139~fb$^{-1}$ of $pp$ collisions at $\sqrt{s}=13$~\TeV\ recorded between 2015 and 2018, the candidate events of interest are required to satisfy either a single-electron trigger~\cite{ATLAS:2019dpa,ATLAS:2021lws} or a single-muon trigger~\cite{ATLAS:2020gty}.  Both triggers had $p_\textrm{T}$ thresholds that increased from 20 to 26~\GeV\ during the data-taking period.
 
Events in the signal region are selected by requiring one electron and one oppositely charged muon with an invariant mass in the window $70 < m_{e\mu} < 110$~\GeV.
Events in the control region are selected by requiring two opposite-charge electrons (muons) with an invariant mass in the window $70 < m_{ee\ (\mu\mu)} < 110$~\GeV, to estimate the  expected number of $Z$ bosons.
Events with more than two candidate leptons are vetoed, using the ``Loose'' electron~\cite{Aad:2019tso} or muon~\cite{Aad:2016jkr} identification criteria. To suppress the top quark background, candidate events are vetoed if they contain a leading jet with $p_{\textrm{T}} > 60$~\GeV, or $E_\textrm{T}^\textrm{miss} > 50$~\GeV, or any jets tagged as containing $b$-hadrons.
 
\begin{figure}[h]
\centering
\includegraphics[width=0.7\textwidth]{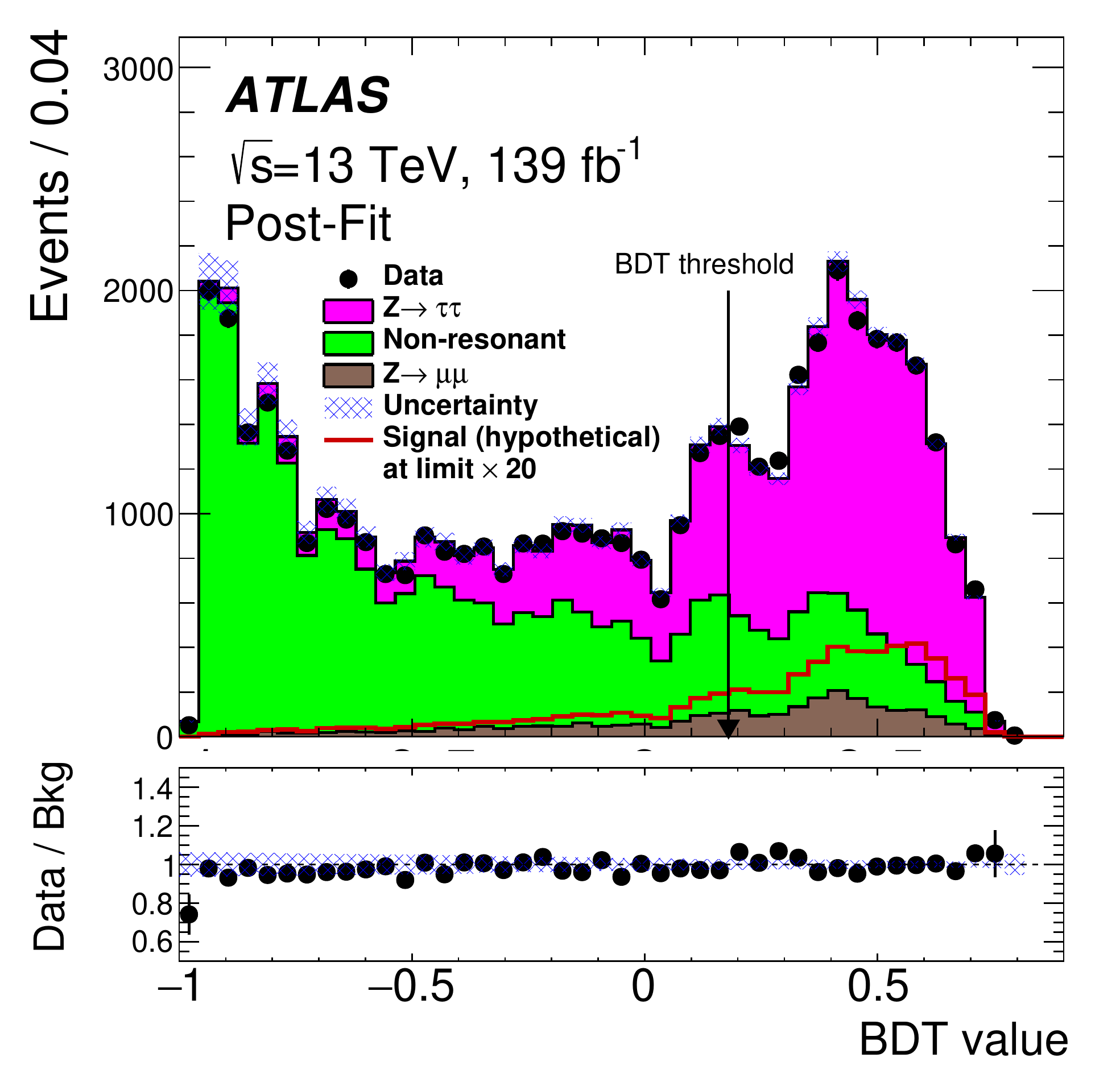}
\caption{
Distributions of BDT values for samples of simulated $Z\rightarrow e\mu$ signal and background events (histograms), as well as data (points).
The component labeled ``non-resonant'' includes all SM background processes except $Z\rightarrow \tau\tau$ and $Z\rightarrow \mu\mu$.
The normalizations for the $Z\rightarrow \tau\tau$, $Z\rightarrow \mu\mu$, and non-resonant background use the best-fit values in Table~\ref{tab:fit}.
A hypothetical $Z\rightarrow e\mu$ signal, whose branching fraction is set to 20 times the observed upper limit, is shown with a dark red solid line for illustration purposes.
The uncertainty band for the MC histogram distribution includes both the systematic and statistical uncertainty. An arrow is added to show the analysis selection requirement for the BDT, corresponding to a value of 0.18.  }
\label{fig:bdt}
\end{figure}

A machine-learning strategy is used to find the optimal selection in the three-dimensional space of leading jet $p_{\textrm{T}}, E_\textrm{T}^\textrm{miss},$ and $p^{e\mu}_{\textrm{T}}$ to further suppress the background and enhance the signal; if no jets are reconstructed, a value of zero is used for the leading jet $p_{\textrm{T}}$. A gradient BDT is trained on samples of simulated signal and background events in the mass window $85 < m_{e\mu} < 95$~\GeV, excluding the $Z\to\mu\mu$ background, where large event weights lead to unstable performance in training. The threshold value for the BDT output is chosen by maximizing the ratio of the  expected signal to the square root of the expected background. Distributions of BDT values for simulated signal and background events in the mass window $70 < m_{e\mu} < 110$~\GeV\  are shown in Figure~\ref{fig:bdt} along with the ratio of data to MC background events as a function of BDT value. There is no evidence of overtraining, as performance is consistent between testing and training samples.

An analogous selection, but for same-flavor lepton pairs, is applied to build the control region sample used to calculate a normalization which eliminates many systematic uncertainties. For the  $ee\ (\mu\mu)$ control region sample, $p^{ee}_{\textrm{T}} (p^{\mu\mu}_{\textrm{T}})$ is used as an input to the BDT.
 
\section{Background estimation}
 
The dominant background in the full mass range considered is due to $Z\rightarrow \tau\tau \rightarrow e \mu \nu_e\nu_\mu\nu_{\tau}\nu_\tau$.
The subleading background is due to $Z\rightarrow \mu\mu$ decays where a muon is misidentified as an electron, due to either muon decay, or radiation of a photon,
or 
an unusually large energy deposit in the EM calorimeter by the muon. Both the $Z\rightarrow \tau\tau$ and $Z\rightarrow \mu\mu$ contributions are modeled using simulated events.
 
Additional backgrounds are due to diboson processes,
top quark single or pair production, and leptonically decaying $W$ bosons. These backgrounds are modeled using samples of simulated events for studies such as validation of the background estimation method, but these samples are not used in the final likelihood fit.
The final background, in which two lepton candidates are misidentified jets, is estimated by extrapolating from samples of data events with leptons of the same electric charge. Assuming that jets are equally likely to be misidentified with either charge, the same-charge contribution is used as an estimate of the opposite-charge contribution, after subtracting previously accounted-for processes, estimated using  samples of simulated events, to avoid double-counting.
The non-resonant backgrounds, i.e. those other than $Z\rightarrow \tau\tau$ and $Z\rightarrow \mu\mu$,  are the dominant background within the narrower mass range where the signal is expected. Roughly 80\% of this non-resonant background comes from diboson processes and 10\% consists of two jets misidentified as lepton candidates.
A second-order polynomial function is used to describe the distribution of the non-resonant backgrounds and also to correct for residual differences between the data and the backgrounds estimated from simulation samples, where the normalization and functional parameters float in the fit.
The functional form was validated against models of these backgrounds which use  samples of simulated events to describe the diboson and top quark contributions and use samples of data and simulated events with leptons of the same electric charge to describe the fake-lepton contributions.

The branching fraction of $Z\rightarrow e\mu$ events is estimated using a binned extended maximum-likelihood fit to the $m_{e\mu}$ distribution, where the likelihood is also a function of the number of $Z\rightarrow\tau\tau$ events, the number of $Z\rightarrow \mu\mu$ events and the number of events due to all non-resonant backgrounds, all of which are free to float in the fit.
The distributions in $m_{e\mu}$ for the $Z\rightarrow e\mu$ signal and the $Z\rightarrow\tau\tau$ and $Z\rightarrow \mu\mu$
backgrounds are modeled using histograms based on the  samples of simulated events.

\section{Results}
\label{sec:result}

\begin{figure}[h]
\centering
\includegraphics[width=0.7\textwidth]{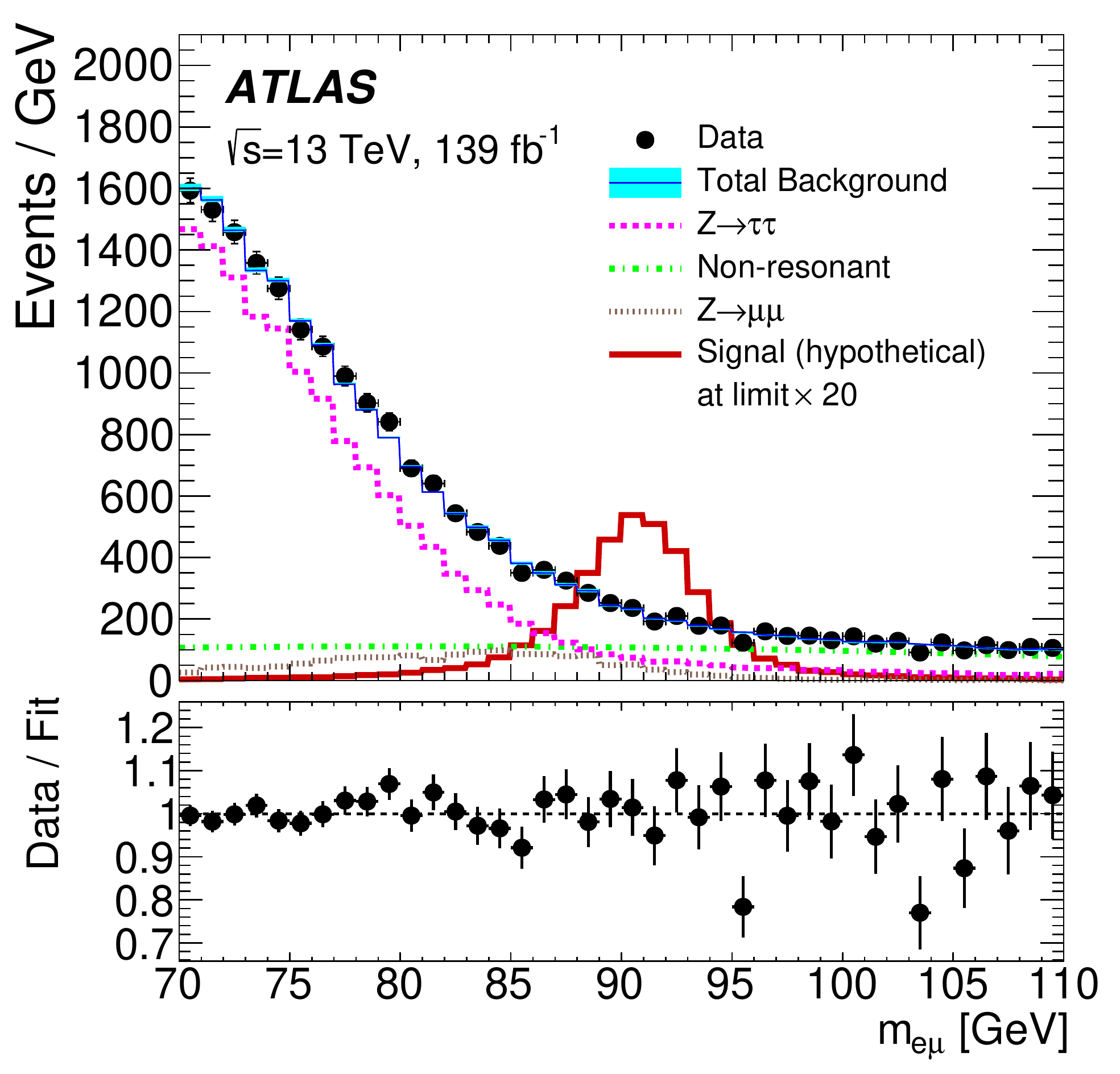}
\caption{ Distribution of the invariant mass $m_{e\mu}$ of the $Z\rightarrow e\mu$ candidates, for data (points) and expected backgrounds (lines) after the background-only likelihood fit. The goodness-of-fit, as measured by the $\chi^2$ divided by the number of degrees of freedom is 1.2, with probability 0.24. The final total fit is shown with a blue solid line, the $Z\to\tau\tau$ component with a green dashed line, the $Z\to\mu\mu$ component with a brown dotted line, and the pink dash-dotted curve represents all non-resonant background contributions. The statistical uncertainty is shown with the light blue band.
A hypothetical $Z\rightarrow e\mu$ signal, its branching fraction scaled to 20 times the observed upper limit, is shown as the dark red solid line for illustration purposes.
The lower panel shows the ratio of observed data to expected background yields. }
\label{fig:sig}
\end{figure}

Distributions of observed events, expected backgrounds after a background-only fit, and a benchmark signal are shown in Figure~\ref{fig:sig}.   The spectrum of $m_{e\mu}$ is consistent with the background expectation, with no evidence of an enhancement near the $Z$ boson mass. 
The best-fit values of the contributions from the various background components are given in Table~\ref{tab:fit}, and are consistent
with the numbers from the simulated events and the estimated fake-lepton contributions.
 
\begin{table}
\caption{ Best-fit values of background contributions for the $Z\to\tau\tau$, $Z\to\mu\mu$ and non-resonant backgrounds, from a background-only fit to data corresponding to 139~fb$^{-1}$, and the expected numbers from the simulated events and the estimated fake-lepton contributions, reported for the full mass window $[70,110]$~\GeV, and for a signal-enriched subwindow $[85,95]$~\GeV. The statistical uncertainty, as defined in the text, is shown.}
\vspace{0.2cm}
\centering
\label{tab:fit}
\begin{tabular}{lcccc}
\hline
\hline
& \multicolumn{2}{c}{Best-fit contribution in mass window} & \multicolumn{2}{c}{Expected contribution in mass window}\\
Background & $[70,110]$~\GeV & $[85,95]$~\GeV & $[70,110]$~\GeV & $[85,95]$~\GeV \\
\hline
$Z\to\tau\tau$ & $13716 \pm 185$~~ &$951 \pm 13$ & $13750$ & $953$ \\
$Z\to\mu\mu$ &  $1557 \pm 209$ & $533 \pm 72$ & $1341$ & $459$ \\
Non-resonant  & $4105 \pm 259$ & $1075 \pm 68$ & $3728$ & $1003$~~ \\
\hline
\hline
\end{tabular}
\end{table}
 
The branching fraction of $Z\rightarrow e\mu$, $\mathcal{B}(Z\rightarrow e\mu)$, is related to the number of $Z\rightarrow e\mu$ decays ($N_{Z\rightarrow e\mu}$) divided by the product of the $Z\rightarrow e\mu$ signal acceptance and efficiency, $(A\times\varepsilon)_{Z\rightarrow e\mu}$, and the number of $Z$ bosons expected in the sample ($N_Z^{\textrm{avg}}$):
 
\begin{equation}
N_{Z\rightarrow e\mu} = N_Z^{\textrm{avg}} \times (A\times\varepsilon)_{Z\rightarrow e\mu} \times  \mathcal{B}(Z\rightarrow e\mu)
\label{eq1}
\end{equation}
 
\noindent
where $N_Z^{\textrm{avg}}$ is the estimate of the number of $Z$ boson events produced, as measured and geometrically averaged from samples of $ee$ and $\mu\mu$ events with invariant mass in the range of $[85,95]$~\GeV, selected with the same requirements as the $e\mu$ sample, other than the same-lepton-flavor requirement, and corrected for background contributions, acceptance times efficiency and the $Z$ leptonic branching ratio. Acceptance and efficiency are measured in  samples of simulated events. Comparisons of jet momentum, pseudorapidity and multiplicity in the $Z\rightarrow e\mu$ sample, simulated at LO, with the respective quantities in the $Z\rightarrow \mu\mu$, simulated at NLO, show negligible differences after reweighting, and affect the $Z\rightarrow e\mu$ branching fraction by less than 0.1\%.
 
The estimate of $\mathcal{B}(Z\rightarrow e\mu)$ is extracted using a maximum-likelihood signal-plus-background fit in which $\mathcal{B}(Z\rightarrow e\mu)$ is the paramter of interest and which incorporates nuisance parameters for the systematic uncertainties and the parameters of the second-order polynomial function used to model the non-resonant background. The best estimate is $\mathcal{B}(Z\rightarrow e\mu) = (0.3 \pm 1.1 (\textrm{stat}) \pm 0.6 (\textrm{syst}))\times 10^{-7}$. The statistical uncertainty is determined by fixing the nuisance parameters to their best-fit values. The systematic uncertainty is determined by subtracting the square of the statistical uncertainty from the square of the total uncertainty. An upper limit on $\mathcal{B}(Z\rightarrow e\mu)$ is calculated at 95\% CL using a one-sided profile-likelihood test statistic in the asymptotic approximation~\cite{Cowan:2010js}.
The values of $\varepsilon_{Z\rightarrow e\mu}$ and $N_Z^{\textrm{avg}}$ are given in Table~\ref{tab:res}. This method gives a limit which is insensitive to sources of systematic uncertainty which are correlated between the $ee$, $\mu\mu$ and $e\mu$ final states, such as the jet \pT threshold efficiency, modeling of $E_\textrm{T}^\textrm{miss}$ in simulation,  and the integrated luminosity.
 
The dominant remaining systematic uncertainties are due to the statistical uncertainty of the simulated events used to form histograms of the $Z\rightarrow\tau\tau$
and $Z\rightarrow \mu\mu$ backgrounds, which are applied independently to each bin, allowing the bin content to vary.
The expected upper limit including these uncertainties is 9.5\% higher than
the limit expected without these uncertainties.
 
Additional uncertainties include the jet energy scale and resolution uncertainty~\cite{PERF-2016-04, PERF-2011-04},
uncertainties in the efficiency of the jet $b$-tagging~\cite{Aad:2019aic}, and the pileup reweighting uncertainties~\cite{STDM-2015-05}. Further sources of systematic uncertainty are the uncertainties in lepton trigger, reconstruction, identification and isolation efficiencies~\cite{PERF-2015-10,PERF-2018-01}.  Electrons have additional uncertainties from energy scale and resolution uncertainties, while muons have uncertainties from momentum resolution, track-to-vertex matching, and sagitta bias correction uncertainties.
These additional systematic uncertainties vary the shape of the histograms for the $Z\rightarrow e\mu$,
$Z\rightarrow\tau\tau$ and $Z\rightarrow \mu\mu$ processes, and change the signal efficiency. The effects of these additional systematic uncertainties degrade the expected limit by 2.4\%. Uncertainties due to higher-order corrections to the simulated $Z\rightarrow e\mu$ signal as well as uncertainties due to potential mismodeling of the simulated background processes are found to be negligible.
 
Any potential bias due to the inability of the chosen polynomial function to accurately describe the background is estimated by fitting the simulated signal-plus-background model to samples of simulated background events. The resulting bias is found to be negligible.
 
Table~\ref{tab:sysbreakdown} shows a summary of the uncertainties and their impact on the expected upper limit of the signal branching fraction $\mathcal{B}$($Z\rightarrow e\mu$). The  observed (expected) upper limit on the $\mathcal{B}$($Z\rightarrow e\mu$) is  $2.62\ (2.37) \times 10^{-7}$ at 95\% CL, which would correspond to approximately 200 $Z\rightarrow e\mu$ reconstructed events.
The larger integrated luminosity (139~fb$^{-1}$) and higher energy ($\sqrt{s}=13$~\TeV) for this search lower the expected upper limit by a factor of three
relative to the previous ATLAS result.
 
\begin{table}
\caption{Values of quantities used to calculate $\mathcal{B}$($Z\rightarrow e\mu$) via Eq.~(\ref{eq1}). Quoted uncertainties reflect statistical and systematic contributions.}
\vspace{0.2cm}
\centering
\label{tab:res}
\begin{tabular}{lr}
\hline
\hline
Quantity & Value \\
\hline
$A\times\varepsilon_{Z\rightarrow e\mu}$ & $(10.3 \pm 0.3)$\% \\
$N_Z^{\textrm{avg}}$ &  $(7.87 \pm 0.19) \times 10^9$\\
\hline
\hline
\end{tabular}
 
\end{table}
 
\begin{table}
\caption{ Effect of various sources of systematic uncertainty on the expected upper limit on the branching fraction $\mathcal{B}$($Z\rightarrow e\mu$), measured by comparing the expected limits obtained with and without a given source of uncertainty.  Uncertainties due to the statistical uncertainty of the samples of simulated events used to form the histograms which describe the $Z\to\tau\tau$ and $Z\to\mu\mu$ background processes are treated as systematic uncertainties.}
\vspace{0.2cm}
\centering
\label{tab:sysbreakdown}
\begin{tabular}{lr}
\hline
\hline
Source of uncertainty & Degradation of $\mathcal{B}^{95\%\textrm{CL}}$($Z\rightarrow e\mu$) \\
\hline
Statistical uncertainty in MC samples & 9.5\% \quad \quad \\
\quad $Z\to\tau\tau$ & \quad 4.7\% \\
\quad $Z\to\mu\mu$   & \quad 6.1\% \\
All other sources            & 2.4\% \quad \quad \\
\quad Jet energy scale and resolution & \quad 1.2\% \\
\quad Pileup        & \quad 1.2\% \\
\quad Electron energy scale and resolution  & \quad 0.8\% \\
\quad Lepton efficiency  & \quad 0.7\% \\
\quad $b$-tagging       & \quad 0.6\% \\
\quad Muon resolution and bias correction  &\quad 0.6\% \\
\hline
\hline
\end{tabular}
 
\end{table}

\FloatBarrier

\section{Conclusion}
 
A search for the lepton-flavor-violating process $Z\rightarrow e\mu$ is performed in 139~fb$^{-1}$ of $\sqrt{s}=13$~\TeV\ proton collision data at collected by the ATLAS experiment at the LHC.  No localized excess consistent with such a decay is observed in the $m_{e\mu}$ spectrum.  An upper limit of $\mathcal{B}$($Z\rightarrow e\mu) < 2.62 \times 10^{-7}$ is set at 95\% CL, a significant improvement on the previous LHC limit, and the most stringent direct result yet reported.

\section*{Acknowledgments}


We thank CERN for the very successful operation of the LHC, as well as the
support staff from our institutions without whom ATLAS could not be
operated efficiently.
 
We acknowledge the support of
ANPCyT, Argentina;
YerPhI, Armenia;
ARC, Australia;
BMWFW and FWF, Austria;
ANAS, Azerbaijan;
CNPq and FAPESP, Brazil;
NSERC, NRC and CFI, Canada;
CERN;
ANID, Chile;
CAS, MOST and NSFC, China;
Minciencias, Colombia;
MEYS CR, Czech Republic;
DNRF and DNSRC, Denmark;
IN2P3-CNRS and CEA-DRF/IRFU, France;
SRNSFG, Georgia;
BMBF, HGF and MPG, Germany;
GSRI, Greece;
RGC and Hong Kong SAR, China;
ISF and Benoziyo Center, Israel;
INFN, Italy;
MEXT and JSPS, Japan;
CNRST, Morocco;
NWO, Netherlands;
RCN, Norway;
MEiN, Poland;
FCT, Portugal;
MNE/IFA, Romania;
MESTD, Serbia;
MSSR, Slovakia;
ARRS and MIZ\v{S}, Slovenia;
DSI/NRF, South Africa;
MICINN, Spain;
SRC and Wallenberg Foundation, Sweden;
SERI, SNSF and Cantons of Bern and Geneva, Switzerland;
MOST, Taiwan;
TENMAK, T\"urkiye;
STFC, United Kingdom;
DOE and NSF, United States of America.
In addition, individual groups and members have received support from
BCKDF, CANARIE, Compute Canada and CRC, Canada;
PRIMUS 21/SCI/017 and UNCE SCI/013, Czech Republic;
COST, ERC, ERDF, Horizon 2020 and Marie Sk{\l}odowska-Curie Actions, European Union;
Investissements d'Avenir Labex, Investissements d'Avenir Idex and ANR, France;
DFG and AvH Foundation, Germany;
Herakleitos, Thales and Aristeia programmes co-financed by EU-ESF and the Greek NSRF, Greece;
BSF-NSF and MINERVA, Israel;
Norwegian Financial Mechanism 2014-2021, Norway;
NCN and NAWA, Poland;
La Caixa Banking Foundation, CERCA Programme Generalitat de Catalunya and PROMETEO and GenT Programmes Generalitat Valenciana, Spain;
G\"{o}ran Gustafssons Stiftelse, Sweden;
The Royal Society and Leverhulme Trust, United Kingdom.
 
The crucial computing support from all WLCG partners is acknowledged gratefully, in particular from CERN, the ATLAS Tier-1 facilities at TRIUMF (Canada), NDGF (Denmark, Norway, Sweden), CC-IN2P3 (France), KIT/GridKA (Germany), INFN-CNAF (Italy), NL-T1 (Netherlands), PIC (Spain), ASGC (Taiwan), RAL (UK) and BNL (USA), the Tier-2 facilities worldwide and large non-WLCG resource providers. Major contributors of computing resources are listed in Ref.~\cite{ATL-SOFT-PUB-2021-003}.


\printbibliography
 
\clearpage
 
\begin{flushleft}
\hypersetup{urlcolor=black}
{\Large The ATLAS Collaboration}

\bigskip

\AtlasOrcid[0000-0002-6665-4934]{G.~Aad}$^\textrm{\scriptsize 99}$,
\AtlasOrcid[0000-0002-5888-2734]{B.~Abbott}$^\textrm{\scriptsize 117}$,
\AtlasOrcid[0000-0002-7248-3203]{D.C.~Abbott}$^\textrm{\scriptsize 100}$,
\AtlasOrcid[0000-0002-2788-3822]{A.~Abed~Abud}$^\textrm{\scriptsize 34}$,
\AtlasOrcid[0000-0002-1002-1652]{K.~Abeling}$^\textrm{\scriptsize 53}$,
\AtlasOrcid[0000-0002-2987-4006]{D.K.~Abhayasinghe}$^\textrm{\scriptsize 92}$,
\AtlasOrcid[0000-0002-8496-9294]{S.H.~Abidi}$^\textrm{\scriptsize 27}$,
\AtlasOrcid[0000-0002-9987-2292]{A.~Aboulhorma}$^\textrm{\scriptsize 33e}$,
\AtlasOrcid[0000-0001-5329-6640]{H.~Abramowicz}$^\textrm{\scriptsize 148}$,
\AtlasOrcid[0000-0002-1599-2896]{H.~Abreu}$^\textrm{\scriptsize 147}$
% ...
\bigskip
\\

% ...
$^{27}$Physics Department, Brookhaven National Laboratory, Upton NY; United States of America.\\
% ...
$^{33}$$^{(a)}$Facult\'e des Sciences Ain Chock, R\'eseau Universitaire de Physique des Hautes Energies - Universit\'e Hassan II, Casablanca;$^{(b)}$Facult\'{e} des Sciences, Universit\'{e} Ibn-Tofail, K\'{e}nitra;$^{(c)}$Facult\'e des Sciences Semlalia, Universit\'e Cadi Ayyad, LPHEA-Marrakech;$^{(d)}$LPMR, Facult\'e des Sciences, Universit\'e Mohamed Premier, Oujda;$^{(e)}$Facult\'e des sciences, Universit\'e Mohammed V, Rabat;$^{(f)}$Institute of Applied Physics, Mohammed VI Polytechnic University, Ben Guerir; Morocco.\\
$^{34}$CERN, Geneva; Switzerland.\\
% ...
$^{53}$II. Physikalisches Institut, Georg-August-Universit\"{a}t G\"ottingen, G\"ottingen; Germany.\\
% ...
$^{92}$Department of Physics, Royal Holloway University of London, Egham; United Kingdom.\\
% ...
$^{99}$CPPM, Aix-Marseille Universit\'e, CNRS/IN2P3, Marseille; France.\\
$^{100}$Department of Physics, University of Massachusetts, Amherst MA; United States of America.\\
% ...
$^{117}$Homer L. Dodge Department of Physics and Astronomy, University of Oklahoma, Norman OK; United States of America.\\
% ...
$^{147}$Department of Physics, Technion, Israel Institute of Technology, Haifa; Israel.\\
$^{148}$Raymond and Beverly Sackler School of Physics and Astronomy, Tel Aviv University, Tel Aviv; Israel.\\
% ...
$^{a}$ Also Affiliated with an institute covered by a cooperation agreement with CERN.\\
$^{b}$ Also at Borough of Manhattan Community College, City University of New York, New York NY; United States of America.\\
$^{c}$ Also at Bruno Kessler Foundation, Trento; Italy.\\
$^{d}$ Also at Center for High Energy Physics, Peking University; China.\\
$^{e}$ Also at Centro Studi e Ricerche Enrico Fermi; Italy.\\
$^{f}$ Also at CERN, Geneva; Switzerland.\\
% ...

\end{flushleft}


\end{document}